\documentclass[12pt,letterpaper]{article}
\usepackage[utf8]{inputenc}
\usepackage{xcolor}
\usepackage{natbib}
\usepackage[colorlinks=true,linkcolor=blue,anchorcolor=blue,citecolor=blue,filecolor=black,menucolor=black,runcolor=black,urlcolor=blue]{hyperref}
\usepackage{url}
\usepackage{graphicx}
\usepackage{float}

\usepackage[symbol]{footmisc}

\usepackage[margin=1.25in,footskip=0.25in]{geometry}

\usepackage{amsmath, amssymb, mathtools, mathrsfs}

\usepackage{setspace}
\newcommand{\orcid}[1]{\href{https://orcid.org/#1}{\includegraphics[width=10pt]{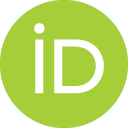}}}

\begin{document}

\section*{\centering \texttt{hankl}: A lightweight Python implementation of the FFTLog algorithm for Cosmology}

\noindent
\begin{center}
\textbf{Minas Karamanis} \orcid{0000-0001-9489-4612}$^{1}$\footnote[2]{E-mail: minas.karamanis@ed.ac.uk}, and  \textbf{Florian Beutler} \orcid{0000-0003-0467-5438}$^{1}$\\
\textsl{\footnotesize $^{1}$Institute for Astronomy, University of Edinburgh, Royal Observatory, Blackford Hill,\\ Edinburgh EH9 3HJ, UK}
\end{center}

\vspace{0.25cm}
\begin{abstract}
    We introduce \texttt{hankl}, a lightweight Python implementation of the FFTLog algorithm for Cosmology. The FFTLog algorithm is an extension of the Fast Fourier Transform (FFT) for logarithmically spaced periodic sequences. It can be used to efficiently compute Hankel transformations, which are paramount for many modern cosmological analyses that are based on the power spectrum or the 2-point correlation function multipoles. The code is well-tested, open source, and publicly available at \url{https://github.com/minaskar/hankl}.
\end{abstract}

\thispagestyle{empty}

\section{Introduction}

The Hankel transform (also known as the Fourier-Bessel transform) is an integral transformation whose kernel is a Bessel function. The Hankel transform appears very often in physical problems with spherical or cylindrical symmetry as it emerges when one writes the usual Fourier transform in spherical coordinates. The Hankel transform finds application in a wide range of scientific fields, namely cosmology, astrophysics, geophysics, and fluid mechanics.

As an example, in modern cosmology, the large-scale clustering of galaxies in the observable universe is often described by means of the configuration-space 2-point correlation function and its Fourier space counterpart, the power spectrum \citep{peebles2020large}. Due to the statistical isotropy of the universe these two quantities are related by a Hankel transformation. In the 3D case, the pair is related by a sine transform which can be considered as a special case of the Hankel transform. The ability to perform such transformations in a fast and accurate manner is of paramount importance for studies of the large-scale structure of the universe.

However, the implementation of the Hankel transform poses some serious numerical challenges. Most importantly, the Bessel function kernel is a highly oscillatory function and any naive implementation of the quadrature numerical integration methods could lead to grossly inaccurate results. To successfully overcome this issue, \citet{talman1978numerical}, and later \citet{hamilton2000uncorrelated}, introduced the FFTLog algorithm, which can be thought of as the Fast Fourier Transform of a logarithmically spaced periodic sequence.

The Hankel transform pair, most commonly appearing in the cosmological literature, has the form~\citep{hamilton2000uncorrelated}
\begin{subequations}
\label{eq:1}
\begin{align}
    f(k)= \int_0^\infty F(x) (kr)^{q} J_\mu(kr) k dr\, ,   \label{eq:1a}\\
    F(r)= \int_0^\infty f(k) (kr)^{-q} J_\mu(kr) r dk\,.   \label{eq:1b}
\end{align}
\end{subequations}
If the substitution $F(r)=g(r)\,r^{-q}$ and $f(k)=G(k)\,y^{q}$ is made, then the aforementioned transform pair reduces to the standard form of the Hankel transform
\begin{subequations}
\label{eq:2}
\begin{align}
    g(k)= \int_0^\infty G(r) J_\mu(kr) k dr\, ,   \label{eq:2a}\\
    G(r)= \int_0^\infty f(k) J_\mu(kr) r dk\,.   \label{eq:2b}
\end{align}
\end{subequations}
In the continuous case, the transform pairs of equations \ref{eq:1} and \ref{eq:2} are equivalent. The transformations are different when they are made discrete in periodic sequences. The FFTLog algorithm computes such discrete transforms with arbitrary power--law bias $(kr)^{\pm q}$ of the form of equations \ref{eq:1}. In terms of the $\ln{r}$ and $\ln{k}$ variables, the above integral transformations become a convolution which can be evaluated as a multiplication in Fourier space. This is the main idea behind Fast Hankel Transforms~\citep{siegman1977quasi} and FFTLog~\citep{talman1978numerical}.

\section{Implementation}

\texttt{hankl} is a lightweight \texttt{Python} implementation of the FFTLog algorithm that particularly focuses on cosmological applications. \texttt{hankl} relies on the \texttt{NumPy} and \texttt{SciPy} libraries in order to provide fast and accurate Hankel transforms with minimal computational overhead. \texttt{hankl} is well suited for scientific applications that require a simple and modular \texttt{Python} interface along with C-level performance.

\texttt{hankl} is designed to perform general Hankel transformations of the form seen in equations \ref{eq:1} as well as the more cosmologically relevant transformations between the power spectrum and the 2-point correlation function, such as
\begin{equation}
    \xi_{\ell}^{(n)}(r) = i^{\ell} \int_{0}^{\infty} k^{2} \frac{dk}{2 \pi^{2}} (kr)^{-n} P_{\ell}^{(n)}(k) j_{\ell}(ks)\, ,
\end{equation}
and the inverse
\begin{equation}
    P_{\ell}^{(n)}(k) = 4 \pi (-i)^{\ell} \int_{0}^{\infty} r^{2} dr (kr)^{n} \xi_{\ell}^{(n)}(r) j_{\ell}(kr)\, .
\end{equation}

\begin{figure}
    \centering
	\centerline{\includegraphics[scale=0.6]{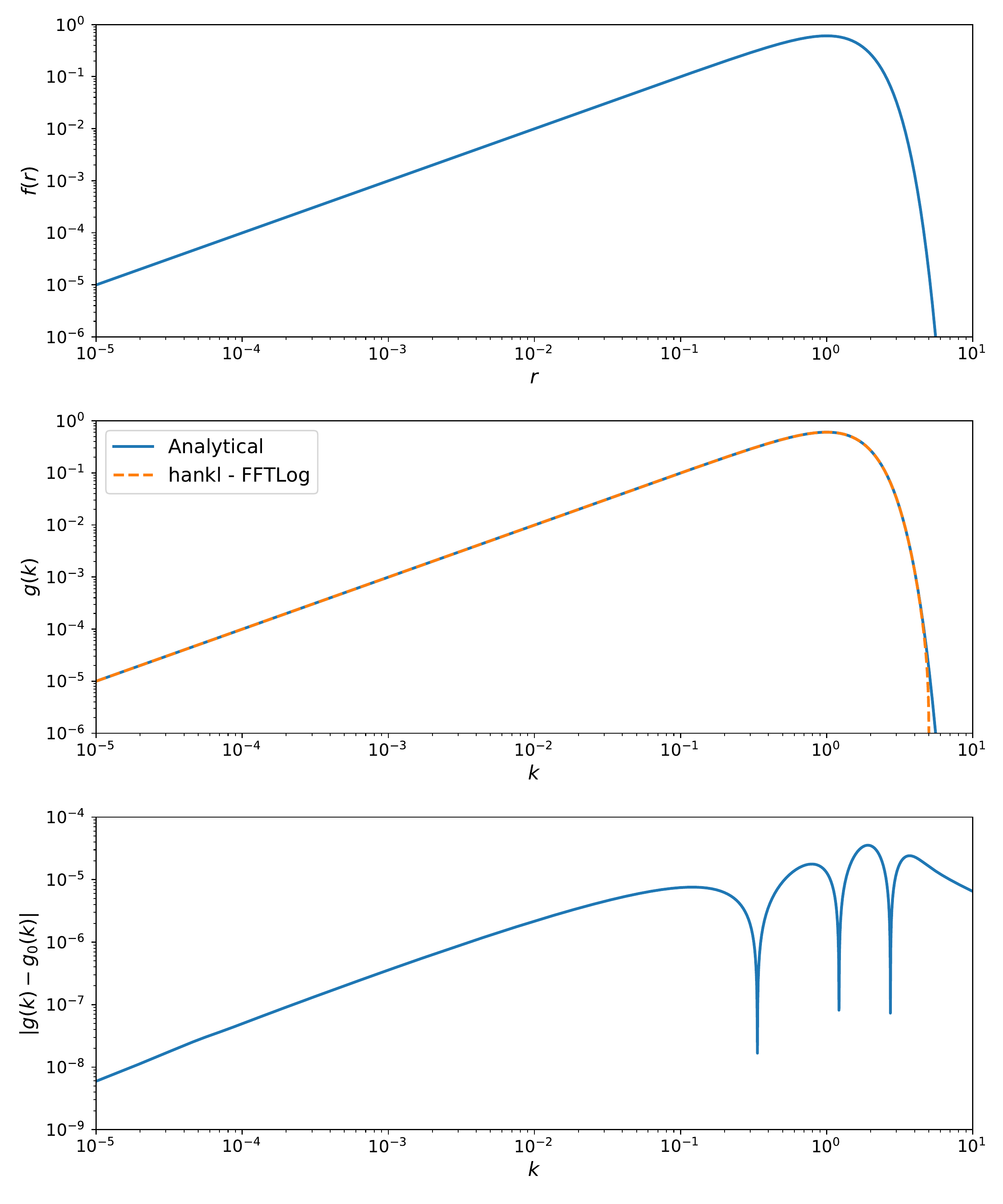}}
    \caption{The plot shows the Hankel transform $g(k)$ (middle panel) of $f(r)$ (top panel) as well as the absolute difference (bottom panel) of the result estimated with FFTLog (orange) and the exact analytical result $g_{0}(k)$ (blue).}
    \label{fig:test}
\end{figure}

One example, with analytically tractable Hankel transformation, is shown in figure \ref{fig:test}. The original function is
\begin{equation}
    \label{eq:f}
    f(r) = r^{\mu+1}\exp{\bigg(-\frac{r^{2}}{2}\bigg)}\, ,
\end{equation}
with its Hankel transform given by
\begin{equation}
    \label{eq:g}
    g(k) = k^{\mu+1}\exp{\bigg(-\frac{k^{2}}{2}\bigg)}\,.
\end{equation}
The figure shows the case of $\mu=0$. The absolute difference between calculated and analytic results is less than $4\times 10^{-5}$ and it can be reduced further, according to the needs of the analysis, by extending the integration range and increasing the sampling resolution.

\begin{figure}
    \centering
	\centerline{\includegraphics[scale=0.6]{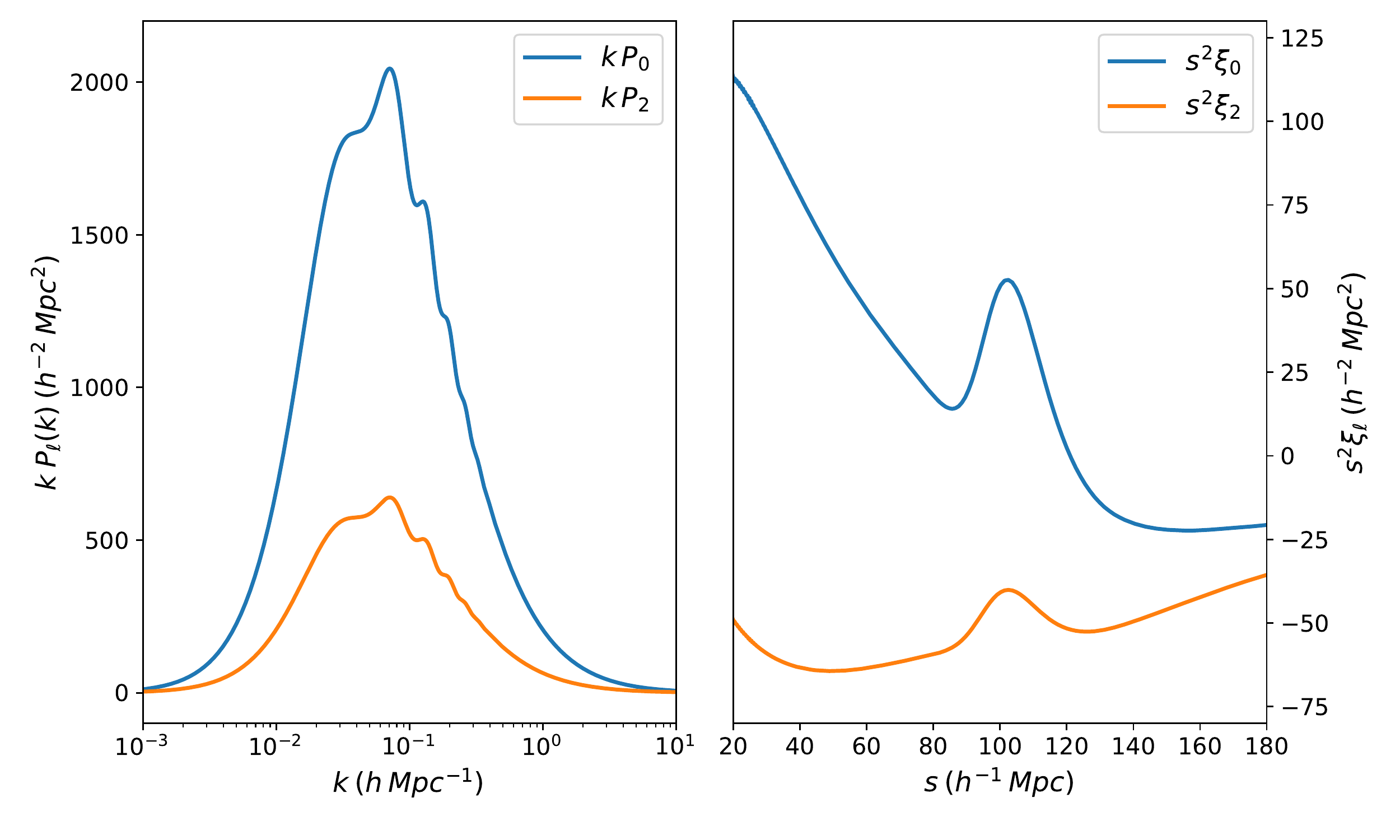}}
    \caption{The plot shows the 2-point correlation function (right panel) monopole $\xi_{0}(s)$ (blue) and quadrupole $\xi_{2}(s)$ (orange) computed using the Hankel transform of the power spectrum (left panel) monopole $P_{0}(k)$ (blue) and quadrupole $P_{2}(k)$ (orange), respectively. }
    \label{fig:cosmo}
\end{figure}

As discussed in the introduction, a very common application of the Hankel transform in cosmology is the transformation of the power spectrum multipoles $P_{\ell}(k)$ from Fourier space to 2-point correlation function multipoles $\xi_{\ell}(s)$ in configuration space. To demonstrate this transformation using \texttt{hankl} we transformed the monopole $P_{0}(k)$ and the quadrupole $P_{2}(k)$ to their respective correlation function multipoles $\xi_{0}$ and $\xi_{2}$. The power spectrum multipoles are computed using the Legendre expansion formula
\begin{equation}
    P_{\ell}(k) = \frac{2\ell + 1}{2} \int_{-1}^{1}d\mu P(k,\mu)\mathcal{L}_{\ell}(\mu)\, ,
\end{equation}
where $\mathcal{L}_{\ell}$ is the Legendre polynomial of order $\ell$, and $\mu$ is the cosine of the angle between the line-of-sight vector and the Fourier mode $\mathbf{k}$. The 2D power spectrum is computed using the Kaiser formula~\citep{kaiser1987clustering}
\begin{equation}
    P(k,\mu ) = (b+f\mu^{2})^{2} P_{m}(k)\, ,
\end{equation}
where $b$ is the linear bias, $f$ is the logarithmic growth rate, and $P_{m}(k)$ is the linear real-space power spectrum computed using CLASS~\citep{blas2011cosmic}. Figure \ref{fig:cosmo} shows the input power spectra and the resulted correlation function multipoles.

\section{Conclusion}

The aim of this project was to develop a tool that could compute the Hankel transform efficiently and accurately as it is required by modern cosmological analyses. To this end we introduced \texttt{hankl}, a lightweight \texttt{Python} implementation of the FFTLog algorithm, that combines a modular and user-friendly \texttt{Python} interface with C-level performance. We hope that \texttt{hankl} will prove useful to the cosmological community by facilitating modern analyses. The code is publicly available at \url{https://github.com/minaskar/hankl} with detailed documentation and examples that can be found at \url{https://hankl.readthedocs.io}.

\section*{Acknowledgements}

We would like to thank John Peacock for providing constructive comments. This project has received funding from the European Research Council (ERC) under the European Union’s Horizon 2020 research and innovation programme (grant agreement 853291). FB is a University Research Fellow.


\bibliographystyle{style}
\bibliography{ref.bib}


\end{document}